\documentclass[notitlepage,nofootinbib,preprintnumbers,amssymb,superscriptaddress,eqsecnum]{revtex4-2}
\usepackage{amsfonts,amssymb,mathtools,graphicx,color,bm}
\definecolor{ultramarine}{rgb}{0.07, 0.04, 0.56}
\definecolor{cadmiumgreen}{rgb}{0.0, 0.42, 0.24}
\definecolor{indigo(dye)}{rgb}{0.0, 0.25, 0.42}
\usepackage[linktocpage=true]{hyperref}
\hypersetup{
colorlinks=true,
citecolor=ultramarine,
linkcolor=cadmiumgreen,
urlcolor=indigo(dye),
}

\usepackage{autobreak}
\newcommand{\fr}[2]{\frac{#1}{#2}}
\newcommand{\D}{{\rm{d}}}
\newcommand{\pa}{\partial}
\newcommand{\ti}{\tilde}
\newcommand{\na}{\nabla}
\newcommand{\bra}[1]{\left( #1 \right)}  
\newcommand{\brb}[1]{\left[ #1 \right]}  
\newcommand{\brc}[1]{\left\{ #1 \right\}}  
\newcommand{\be}{\begin{equation}}  
\newcommand{\ee}{\end{equation}}
\newcommand{\bem}{\begin{bmatrix}}
\newcommand{\eem}{\end{bmatrix}}
\newcommand{\Mpl}{M_{\rm Pl}}

\newcommand{\ga}{\gamma}

\newcommand{\la}{\lambda}

\newcommand{\mn}{{\mu \nu}}
\newcommand{\mC}{\mathcal{C}}
\newcommand{\mE}{\mathcal{E}}

\newcommand{\mH}{\mathcal{H}}
\newcommand{\mL}{\mathcal{L}}

\begin{document}

\preprint{YITP-21-104, IPMU21-0061}

\title{Nonlinear definition of the shadowy mode in higher-order scalar-tensor theories}

\author{Antonio De Felice}
\affiliation{Center for Gravitational Physics, Yukawa Institute for Theoretical Physics, Kyoto University, Kyoto 606-8502, Japan}

\author{Shinji Mukohyama}
\affiliation{Center for Gravitational Physics, Yukawa Institute for Theoretical Physics, Kyoto University, Kyoto 606-8502, Japan}
\affiliation{Kavli Institute for the Physics and Mathematics of the Universe (WPI), The University of Tokyo, 277-8583, Chiba, Japan}

\author{Kazufumi Takahashi}
\affiliation{Center for Gravitational Physics, Yukawa Institute for Theoretical Physics, Kyoto University, Kyoto 606-8502, Japan}

\begin{abstract}
We study U-DHOST theories, i.e., higher-order scalar-tensor theories which are degenerate only in the unitary gauge and yield an apparently unstable extra mode in a generic coordinate system.
We show that the extra mode satisfies a three-dimensional elliptic differential equation on a spacelike hypersurface, and hence it does not propagate.
We clarify how to treat this ``shadowy'' mode at both the linear and the nonlinear levels.
\end{abstract}

\maketitle

\section{Introduction}\label{sec:intro}

Scalar-tensor theories provide a useful description of modified gravity theories.
When constructing a covariant action containing a metric and a scalar field, one requires the absence of Ostrogradsky ghosts~\cite{Woodard:2015zca,Motohashi:2020psc}, which are associated with higher-order equations of motion~(EOMs).
This requirement poses the so-called degeneracy conditions~\cite{Motohashi:2014opa,Langlois:2015cwa,Motohashi:2016ftl,Klein:2016aiq,Motohashi:2017eya,Motohashi:2018pxg}, which severely constrain the form of the higher-derivative interactions.
Scalar-tensor theories with higher derivatives satisfying the degeneracy conditions are called degenerate higher-order scalar-tensor~(DHOST) theories.
Many classes of DHOST theories have been constructed so far~\cite{Langlois:2015cwa,Crisostomi:2016czh,BenAchour:2016fzp,Takahashi:2017pje,Langlois:2018jdg}, which include the Horndeski~\cite{Horndeski:1974wa,Deffayet:2011gz,Kobayashi:2011nu} and Gleyzes-Langlois-Piazza-Vernizzi theories~\cite{Gleyzes:2014dya} as limiting cases (see \cite{Langlois:2018dxi,Kobayashi:2019hrl} for reviews).

In the context of cosmology, it is natural to assume that the scalar field has a timelike gradient, and hence it is useful to take the so-called unitary gauge where the scalar field is fixed to be a given function of time.
In such a particular choice of coordinate system, one obtains a weaker degeneracy condition, which defines a broader class of theories than the DHOST class.
In particular, theories that are degenerate {\it only} in the unitary gauge are called U-DHOST theories~\cite{DeFelice:2018ewo}.\footnote{Under the unitary gauge, the U-DHOST theories reduce to spatially covariant gravity~\cite{Gao:2014soa,Gao:2014fra,Gao:2018znj}. Conversely, restoring the time diffeomorphism invariance by introducing a St{\"u}ckelberg scalar field, models of spatially covariant gravity are mapped into the union of the DHOST class and the U-DHOST class.}
As such, the U-DHOST theories seem to contain an extra mode originating from higher-order EOMs in a generic coordinate system where the scalar field has an inhomogeneous profile, which apparently signals the Ostrogradsky instability.
However, it was pointed out that the extra mode actually does not propagate, and hence the U-DHOST theories are free of the apparent instability.\footnote{This is not the case when the gradient of the scalar field is not timelike. When the scalar field has a spacelike gradient at least in a part of the spacetime domain of our interest, the Ostrogradsky instability may occur and/or the theory can no longer be trusted as a valid effective field theory, i.e., the knowledge of its ultra-violet continuation is necessary. Hence, the U-DHOST theories are valid only when they are used to describe a spacetime accompanied by the scalar field with a timelike gradient.}
Due to this nature, the extra mode was called a ``shadowy'' mode in \cite{DeFelice:2018ewo}.
Note that the shadowy mode is defined in a coordinate-independent manner as a mode living on a spacelike hypersurface (exactly like the shadows produced by the people on the beach).
Such a non-propagating extra mode was found earlier in the context of massive gravity~\cite{Gabadadze:2004iv} and the khronometric theories~\cite{Blas:2010hb,Blas:2011ni}. 
In these works, the extra mode was said to be ``instantaneous'' as it can be interpreted to propagate with an infinite speed.
It is interesting to note that the U-DHOST theories include the khronometric theories, and hence the shadowy mode is understood as generalization of the instantaneous mode.

In order to elucidate the nature of the shadowy mode, the authors of \cite{DeFelice:2018ewo} studied the following simple toy model in a flat two-dimensional spacetime:
    \be
    L=-\fr{1}{2}X+L_{\rm U}, \qquad
    L_{\rm U}\coloneqq -\fr{\xi}{4}\brb{X\pa^\mu X\pa_\mu X-(\pa^\mu\phi\pa_\mu X)^2}, \label{toy}
    \ee
where $X\coloneqq \pa_\mu\phi\pa^\mu\phi$ and $\xi$ is a constant.
Here, the first term in the Lagrangian is nothing but the canonical kinetic term of the scalar field and note that the second term~$L_{\rm U}$ is vanishing when the scalar field is homogeneous.
The authors of \cite{DeFelice:2018ewo} investigated perturbations about some background solution in two different coordinate systems.
The first one is such that the background field is homogeneous, while in the second one the background field has an inhomogeneous profile but still has a timelike gradient.
For perturbations about a homogeneous background, no higher time derivatives appear in the EOM.
On the other hand, once we consider perturbations about an inhomogeneous background, the EOM contains a fourth time derivative, which apparently signals the existence of an unstable extra mode.
This situation is similar to what happens in the U-DHOST theories.
The point is that, for this toy model, the dangerous extra mode satisfies a one-dimensional differential equation on a spacelike line.
This means that the configuration of the extra mode is completely determined once an appropriate boundary condition is imposed on a spatial boundary, and hence the apparent instability is actually not a problem.

Although the above discussion based on the toy model~\eqref{toy} helps us to infer the nature of the shadowy mode in the U-DHOST theories, it is still nontrivial to predict how the situation changes when the dynamics of the metric is taken into account.
Also, it remains unclear how the shadowy mode shows up at the nonlinear level.
Actually, the instantaneous mode in the khronometric theories was studied only at the level of linear perturbations~\cite{Blas:2010hb,Blas:2011ni}.
The aim of the present paper is to clarify how to understand and deal with the shadowy mode in the presence of gravity even at the nonlinear level.
For demonstration purposes, we focus on one of the simplest examples of the U-DHOST theories and study the effects of the shadowy mode at both the linear and nonlinear levels.
We argue that the existence of the shadowy mode indeed affects the structure of the EOMs, where in particular the Hamiltonian constraint takes the form of a (quasi-)linear partial differential equation, whose principal part has the Laplacian operator defined on a three-dimensional hypersurface.
This is in sharp contrast to the case of DHOST theories where the Hamiltonian constraint fixes the lapse function algebraically in the unitary gauge.
The appearance of such higher spatial derivatives in the U-DHOST theories is indeed due to the existence of the shadowy mode.

The rest of this paper is organized as follows.
We first introduce our model in \S\ref{sec:model}.
In \S\ref{sec:cosmo}, we investigate cosmological perturbations in the U-DHOST theories.
In \S\ref{sec:nonlinear_Hamiltonian}, we perform a nonlinear Hamiltonian analysis for the U-DHOST theories.
The results of these two sections clarify how the shadowy mode appears at the linear and nonlinear levels, respectively.
Then, we discuss how to treat the shadowy mode in \S\ref{sec:shadowy}.
Finally, we draw our conclusions in \S\ref{sec:conclusions}.

\section{The model}\label{sec:model}

We study higher-order scalar-tensor theories described by the following action:
	\be
	S=\int \D^4x\sqrt{-g}\brb{P(X)+Q(X)\Box \phi+F(X)R+\sum_{I=1}^{5}A_I(X)L_I^{(2)}}, \label{U-DHOST}
	\ee
where $P$, $Q$, $F$, and $A_I$ ($I=1,\cdots,5$) are functions of $X\coloneqq \phi_\mu\phi^\mu$ and 
	\be
	L_1^{(2)}\coloneqq \phi^{\mn}\phi_{\mn}, \quad
	L_2^{(2)}\coloneqq (\Box\phi)^2, \quad
	L_3^{(2)}\coloneqq \phi^\mu\phi_{\mn}\phi^\nu\Box\phi, \quad
	L_4^{(2)}\coloneqq \phi^\mu\phi_{\mn}\phi^{\nu\la}\phi_\la, \quad
	L_5^{(2)}\coloneqq (\phi^\mu\phi_{\mn}\phi^\nu)^2,
	\ee 
with $\phi_\mu\coloneqq \na_\mu\phi$ and $\phi_{\mn}\coloneqq \na_\mu\na_\nu\phi$.
Equation~\eqref{U-DHOST} is the most general action up to quadratic order in the second derivative of $\phi$.
For a generic choice of the functions~$F$ and $A_I$, we have higher-order EOMs and, as such, there exists an associated Ostrogradsky ghost.
In order to avoid this problem, it is in general necessary to impose the degeneracy conditions on these functions.
DHOST theories are those satisfying the degeneracy requirement in an arbitrary coordinate system, where the coefficient functions satisfy the following set of conditions~\cite{Langlois:2015cwa}:
	\be
	\begin{split}
	A_2&=-A_1, \\
	A_4&=\fr{1}{8(F-XA_1)^2}\bigl\{4F\brb{3(A_1-2F_{X})^2-2A_3F}-A_3X^2(16A_1F_{X}+A_3F) \\
	&~~~~~~~~~~~~~~~~~~~~~~~~+4X\bra{3A_1A_3F+16A_1^2F_{X}-16A_1F_{X}^2-4A_1^3+2A_3FF_{X}}\bigr\}, \\
	A_5&=\fr{1}{8(F-XA_1)^2}(2A_1-XA_3-4F_{X})\brb{A_1(2A_1+3XA_3-4F_{X})-4A_3F}.
	\end{split} \label{DC}
	\ee
If one requires the degeneracy in a specific coordinate system, a weaker condition is obtained.
In particular, the degeneracy condition in the unitary gauge was derived in \cite{Langlois:2015cwa,DeFelice:2018ewo}, which reads
    \be
    4\brb{X(A_1+3A_2)+2F}\brb{A_1+A_2+X(A_3+A_4)+X^2A_5}=3X(2A_2+XA_3+4F_{X})^2. \label{DCU}
    \ee
Indeed, one can check that \eqref{DCU} is satisfied if the set of degeneracy conditions~\eqref{DC} for the DHOST theories is satisfied.
As mentioned earlier, the U-DHOST theories are higher-order scalar-tensor theories that satisfy the degeneracy conditions {\it only} in the unitary gauge.
Namely, within the class of theories described by the action \eqref{U-DHOST}, the U-DHOST theories are defined as those satisfying \eqref{DCU} but not \eqref{DC}.
The DHOST and U-DHOST theories can be generalized to actions with arbitrary powers of $\phi_\mn$~\cite{DeFelice:2018ewo}, though we restrict ourselves to the action~\eqref{U-DHOST} for simplicity.

In the present paper, for demonstration purposes, we study one of the simplest examples of the U-DHOST theories, the one defined by
    \be
    F(X)=\fr{\Mpl^2}{2}, \qquad
    A_1(X)=A_2(X)=A_3(X)=0, \qquad
    A_4(X)=-XA_5(X)=-\xi X,
    \ee
with $\xi$ being a constant, namely,
    \be
    S=\int \D^4x\sqrt{-g}\brc{\fr{\Mpl^2}{2}R+P(X)+Q(X)\Box \phi 
    -\xi\brb{X\phi^\mu\phi_{\mn}\phi^{\nu\la}\phi_\la-(\phi^\mu\phi_{\mn}\phi^\nu)^2}}. \label{model}
    \ee
The term with $\xi$ here is nothing but the four-dimensional covariant version of $L_{\rm U}$ in equation \eqref{toy} and it characterizes the deviation from the DHOST theories.
Also, we are interested in generic cases where there exists one propagating scalar mode in addition to the two tensor modes in the gravity sector.
Namely, we do not discuss special cases with only two propagating degrees of freedom (plus a non-propagating shadowy mode), e.g., the cuscuton model~\cite{Afshordi:2006ad} (and its extension~\cite{Iyonaga:2018vnu,Iyonaga:2020bmm}) or minimally modified gravity~\cite{DeFelice_2015,Lin:2017oow,Mukohyama:2019unx,Aoki:2018zcv,DeFelice:2020eju,DeFelice:2020prd}, for which the discussion below should be modified. Nevertheless, we expect that our analysis can be generalized to include such cases.

In what follows, we study how the shadowy mode shows up at both the linear and nonlinear levels based on the above simple model of the U-DHOST theories.
We first investigate how the shadowy mode affects cosmological perturbations in the next section.

\section{Shadowy mode in cosmological perturbations}\label{sec:cosmo}

In this section, we study perturbations about homogeneous and isotropic solutions in the U-DHOST theories, which helps us understand the effects of the shadowy mode in the presence of gravity.

\subsection{Cosmological scalar perturbations}

Let us study homogeneous and isotropic solutions where the metric and the scalar field have the following form:
    \be
    g_\mn^{(0)} \D x^\mu \D x^\nu=-N^2(t)\D t^2+a^2(t)\delta_{ij}\D x^i\D x^j, \qquad
    \phi^{(0)}=\phi(t). \label{cosmo_BG}
    \ee
We substitute this ansatz into equation \eqref{U-DHOST} and derive the action written in terms of $N$, $a$, and $\phi$, from which the background EOMs are obtained.
Note that the EOM for $\phi$ is redundant due to the Noether identity associated with the time reparametrization symmetry~\cite{Motohashi:2016prk}.
Hence, we need only the EOMs for $N$ and $a$, which are summarized as follows:
    \be
    \begin{split}
    3\Mpl^2H^2+P-2XP_X+6H\fr{\dot{\phi}}{N}XQ_X&=0, \\
    \Mpl^2\bra{3H^2+2\fr{\dot{H}}{N}}+P+\fr{\dot{\phi}\dot{X}}{N^2}Q_X&=0,
    \end{split}
    \ee
where a dot denotes the time derivative and $H\coloneqq \dot{a}/(Na)$.
Note also that we have $X=-\dot{\phi}^2/N^2$ on the background~\eqref{cosmo_BG}.
These EOMs will be used to simplify the coefficients in the quadratic Lagrangian for cosmological perturbations.
Note that the term with $\xi$ in the Lagrangian~\eqref{model} is vanishing at the background level.\footnote{If we choose a coordinate system where the scalar field has an inhomogeneous profile, the effects of $\xi$ should appear even at the background level. However, in this section, we focus on linear perturbations about the homogeneous background and the effects of $\xi$ on them.}

Let us now study scalar perturbations about the above background, by writing down the metric tensor as follows:
    \be
    \begin{split}
    g_\mn \D x^\mu \D x^\nu&=-N^2(1+2\alpha)\D t^2+2N\pa_i\chi \D t\D x^i+\brb{a^2(1+2\zeta)\delta_{ij}+\bra{\pa_i\pa_j-\fr{\delta_{ij}}{3}\pa^{2}}E}\D x^i\D x^j, \\
    \phi&=\phi^{(0)}+\delta\phi,
    \end{split}
    \ee
with $\pa^{2}\coloneqq \delta^{ij}\pa_i\pa_j$ and the perturbation variables being denoted by $\alpha$, $\chi$, $\zeta$, $E$, and $\delta\phi$. 
We fix the gauge degrees of freedom by setting $E=\delta\phi=0$, which is a complete gauge fixing and hence can be imposed at the Lagrangian level~\cite{Motohashi:2016prk}.
Then, the quadratic Lagrangian takes the form,
    \be
    \mL=Na^3\brb{-3\Mpl^2\fr{\dot{\zeta}^2}{N^2}-\fr{\Mpl^2}{a^2}\zeta\pa^{2}\zeta+\alpha\bra{\Sigma-\xi\fr{\dot{\phi}^6}{N^6}\fr{\pa^{2}}{a^2}}\alpha-\fr{2\Theta}{a^2}\alpha\pa^{2}\chi+6\Theta\alpha\fr{\dot{\zeta}}{N}-\fr{2\Mpl^2}{a^2}\alpha\pa^{2}\zeta+\fr{2\Mpl^2}{a^2}\fr{\dot{\zeta}}{N}\pa^{2}\chi}. \label{qLag}
    \ee
Here, we have defined the following quantities:
    \be
    \Sigma\coloneqq -3\Mpl^2H^2+XP_X+2X^2P_{XX}-6H\fr{\dot{\phi}}{N}\bra{2XQ_X+X^2Q_{XX}}, \qquad
    \Theta\coloneqq \Mpl^2H-\fr{\dot{\phi}^3}{N^3}Q_X.
    \ee
As we shall see below, the effects of the shadowy mode can be better understood in the Hamiltonian language.
The Hamiltonian analysis is also useful to define the shadowy mode in a nonlinear manner in \S\ref{sec:nonlinear_Hamiltonian}.

\subsection{Hamiltonian for scalar perturbations}
\label{ssec:Hamiltonian_cosmo_pert}

From the quadratic Lagrangian~\eqref{qLag}, one can construct the canonical momenta for each perturbation variable as
    \be
    p_\alpha\coloneqq \fr{\pa\mL}{\pa\dot{\alpha}}=0, \qquad
    p_\chi\coloneqq \fr{\pa\mL}{\pa\dot{\chi}}=0, \qquad
    p_\zeta\coloneqq \fr{\pa\mL}{\pa\dot{\zeta}}
    =a^3\bra{-6\Mpl^2\fr{\dot{\zeta}}{N}+6\Theta\alpha+\fr{2\Mpl^2}{a^2}\pa^2\chi}.
    \ee
Hence, we have
    \begin{equation}
    \dot{\zeta}=N\bra{\fr{\Theta}{\Mpl^2}\alpha+\frac{\pa^{2}\chi}{3a^2}-\frac{p_{\zeta}}{6a^{3}\Mpl^{2}}}, \label{eq:dot_zet_p_zet}
    \end{equation}
whereas $p_\alpha$ and $p_\chi$ are subject to the primary constraints~$p_{\alpha}\approx 0$ and $p_{\chi}\approx 0$.
Then, performing the needed Legendre transformation, we find the total Hamiltonian, in which both the primary constraints are imposed by use of Lagrange multipliers~\cite{Henneaux:1992ig,Dirac}, as
    \begin{align}
    H_T=&\;\int \D ^3x\bra{\mH+\la_1p_\alpha+\la_2p_\chi}, \\
    \mH\coloneqq&\;
    \Mpl^2Na\zeta\pa^{2}\zeta-Na^3\alpha\bra{\Sigma+\fr{3\Theta^2}{\Mpl^2}-\xi\fr{\dot{\phi}^6}{N^6}\fr{\pa^{2}}{a^2}}\alpha+2\Mpl^2Na\alpha\pa^{2}\zeta-\fr{\Mpl^2}{3}\fr{N}{a}(\pa^{2}\chi)^2-\fr{N}{12\Mpl^2a^3}p_\zeta^2 \nonumber \\
    &\;+\fr{N\Theta}{\Mpl^2}\alpha p_\zeta+\fr{N}{3a^2}p_\zeta\pa^{2}\chi.
    \end{align}

Next, we require that the primary constraints are conserved under time evolution, which poses secondary constraints.
In doing so, it is useful to define smeared constraints by use of a test function~$\varphi$ and compute their time evolution.
The results are summarized as
    \be
    \begin{split}
    \left\{ \int \D^{3}x\, \varphi p_{\alpha},H_{T}\right\}_{\rm P} \approx \int \D^{3}x\, \varphi \mathcal{C}_{\alpha}\approx 0, \qquad
    \left\{ \int \D^{3}x\, \varphi p_{\chi},H_{T}\right\}_{\rm P} \approx \int \D^{3}x\, \varphi \mathcal{C}_{\chi}\approx 0,
    \end{split}
    \ee
where
    \be
    \mC_\alpha\coloneqq N\brb{2a^3\bra{\Sigma+\fr{3\Theta^2}{\Mpl^2}-\xi\fr{\dot{\phi}^6}{N^6}\fr{\pa^{2}}{a^2}}\alpha-2\Mpl^2a\pa^{2}\zeta-\fr{\Theta}{\Mpl^2}p_\zeta}, \qquad
    \mathcal{C}_{\chi}\coloneqq N\pa^{2} \bra{\frac{2\Mpl^{2}}{3}\pa^{2}\chi-\frac{p_{\zeta}}{3a}}.
    \ee
One can verify that there is no tertiary constraint and all the four constraints ($p_{\alpha}$, $p_{\chi}$, $\mC_{\alpha}$, $\mC_{\chi}$) obtained so far are of second class.
Hence, the number of degrees of freedom of the system in the scalar sector can be computed as follows:
    \begin{align}
    &\fr{1}{2}\brb{\bra{\text{$\#$ of phase-space variables}}
    -2\times \bra{\text{$\#$ of first-class constraints}}
    -\bra{\text{$\#$ of second-class constraints}}} \nonumber \\
    &\qquad =\fr{1}{2}\bra{6-2\times 0-4}=1,
    \end{align}
which is the same number as in the case of the DHOST theories, as expected.

Having specified all the constraints, let us now study the structure of them, in particular, the secondary constraint~$\mathcal{C}_{\alpha}\approx 0$.
This constraint can be rewritten in the form,
    \be
    \hat{L}\alpha\coloneqq 2a\xi\fr{\dot{\phi}^6}{N^6}\pa^{2}\alpha-2a^3\bra{\Sigma+\fr{3\Theta^2}{\Mpl^2}}\alpha
    \approx -2\Mpl^2a\pa^{2}\zeta-\fr{\Theta}{\Mpl^2}p_\zeta. \label{linear-Poisson}
    \ee
It should be noted that this constraint amounts to the Hamiltonian constraint.
Indeed, in the Lagrangian formalism, $\mathcal{C}_{\alpha}=0$ is obtained as the equation of motion for the lapse perturbation~$\alpha$.
When $\xi=0$, i.e., in the case of the DHOST theories, this equation generically fixes $\alpha$ algebraically.
On the other hand, when $\xi\ne 0$, the constraint~\eqref{linear-Poisson} has the form of a Poisson equation, which is a manifestation of the existence of the shadowy mode.
In this case, we put the general solution in the form,
    \begin{equation}
    \alpha=\alpha_{\rm hom}+\alpha_{\rm part},
    \end{equation}
where $\alpha_{\rm part}$ is a particular solution satisfying $\mathcal{C}_{\alpha}\approx 0$. 
Then, the homogeneous part~$\alpha_{\rm hom}$ is obtained as a solution to the following elliptic differential equation:
    \be
    \hat{L}\alpha=\left[2a\xi\fr{\dot{\phi}^6}{N^6}\pa^{2}-2a^3\bra{\Sigma+\fr{3\Theta^2}{\Mpl^2}}\right]\alpha=0, 
    \ee
which fixes $\alpha$ once an appropriate boundary condition is imposed.
As such, $\alpha_{\rm hom}$ represents the shadowy mode.
It is also possible to construct a gauge-invariant quantity that reduces to $\alpha$ in the unitary gauge,
    \begin{equation}
    \alpha_{\rm GI}\coloneqq\alpha-\fr{1}{N}\frac{\partial}{\partial t}\bra{N\frac{\delta\phi}{\dot{\phi}}},
    \end{equation}
which provides a gauge-independent definition of the shadowy mode at the linear level.

\section{Nonlinear definition of the shadowy mode}\label{sec:nonlinear_Hamiltonian}

In this section, we perform a nonlinear Hamiltonian analysis of the U-DHOST theories, which provides a nonlinear definition of the shadowy mode.
We assume that the scalar field has a timelike gradient so that we can take the unitary gauge, but the background spacetime remains arbitrary.

In terms of the Arnowitt-Deser-Misner~(ADM) variables, the metric takes the form,
    \be
    g_\mn \D x^\mu \D x^\nu=-N^2\D t^2+\ga_{ij}(\D x^i+N^i\D t)(\D x^j+N^j\D t),
    \ee
where $N$ is the lapse function, $N^i$ is the shift vector, and $\ga_{ij}$ is the induced metric.
We denote the unit vector normal to a constant-$t$ hypersurface and the projection tensor as 
    \be
    n_\mu\coloneqq -N\delta^0_\mu, \qquad
    h_\mn\coloneqq g_\mn +n_\mu n_\nu. \label{projectiontensor}
    \ee
The extrinsic curvature and the acceleration vector are defined by
    \be
    K_{\mu\nu}\coloneqq h_\mu{}^\alpha \na_\alpha n_\nu, \qquad
    a_\mu\coloneqq n^\alpha \na_\alpha n_\mu,
    \ee
which can be written in terms of the ADM variables as
    \be
    K_{ij}=\fr{1}{2N}\bra{\dot{\ga}_{ij}-D_iN_j-D_jN_i}, \qquad
    a_i=\fr{1}{N}D_iN,
    \ee
with $D_i$ being the covariant derivative associated with $\ga_{ij}$.
The following relations are also useful for translating the covariant Lagrangian into the ADM language under the unitary gauge~\cite{Gleyzes:2013ooa}:
    \be
    \phi_\mu=-\fr{\dot{\phi}}{N}n_\mu, \qquad
    \phi_{\mu\nu}=-\fr{\dot{\phi}}{N}(K_{\mu\nu}-n_\mu a_\nu-n_\nu a_\mu)-\fr{N}{2\dot{\phi}}(n^\alpha \pa_\alpha X)n_\mu n_\nu.
    \ee
Note also that $X=\phi_\mu\phi^\mu=-\dot{\phi}^2/N^2$.

With the above ADM variables, the Lagrangian density in \eqref{U-DHOST} can be written as
    \be
    \mL=N\sqrt{\ga}\brb{\fr{\Mpl^2}{2}\bra{R^{(3)}+K_{ij}K^{ij}-K^2}+P+\ti{Q}K+\xi\fr{\dot{\phi}^6}{N^6}\fr{D_iND^iN}{N^2}}, \label{lagADM}
    \ee
where the function~$\ti{Q}(X)$ is defined so that $\ti{Q}_X=-(-X)^{-1/2}Q_X$.
Note also that $X$ appearing here should be understood as a function of $N$ through $X=-\dot{\phi}^2/N^2$.
The canonical momenta conjugate to $N$, $N^i$, and $\ga_{ij}$ are denoted as $\pi_N$, $\pi_i$, and $\pi^{ij}$, respectively.
Since the action does not contain time derivatives of $N$ and $N^i$, we have $\pi_N=\pi_i=0$.
The canonical momentum~$\pi^{ij}$ can be computed as follows:
    \be
    \pi^{ij}=\fr{\Mpl^2}{2}\sqrt{\ga}\bra{K^{ij}-K\ga^{ij}}+\fr{1}{2}\sqrt{\ga}\ti{Q}\ga^{ij},
    \ee
which can be solved for $K_{ij}$ as
    \be
    K_{ij}=\Mpl^{-2}\brb{\fr{1}{\sqrt{\ga}}(2\pi_{ij}-\pi\ga_{ij})+\fr{\ti{Q}}{2}\ga_{ij}}, \label{ext_curv}
    \ee
with $\pi\coloneqq \ga_{ij}\pi^{ij}$.
Then, the total Hamiltonian is given by
    \be
    H_T=H+\int \D^3x\bra{u_N\pi_N+u^i\pi_i}, \qquad
    H\coloneqq \int \D^3x\bra{\pi^{ij}\dot{\ga}_{ij}-\mL}.
    \ee
More explicitly, $H$ can be written as
    \be
    H=\int \D^3x\bra{\mH_N+N^i\mH_i-\sqrt{\ga}\,\xi\fr{\dot{\phi}^6}{N^6}\fr{D_iND^iN}{N}},
    \ee
where we have defined
    \be
    \mH_N\coloneqq N\sqrt{\ga}\brb{\fr{\Mpl^2}{2}\bra{-R^{(3)}+K_{ij}K^{ij}-K^2}-P}, \qquad
    \mH_i\coloneqq -2D^j\fr{\pi_{ij}}{\sqrt{\ga}}.
    \ee
Here, $K_{ij}$ is regarded as a function of canonical variables through equation \eqref{ext_curv} and $\mH_N$ is no longer linear in $N$.
Then, the Hamiltonian constraint, which corresponds to the consistency condition for the primary constraint~$\pi_N\approx 0$, reads
    \be
    \fr{\pa\mH_N}{\pa N}+\sqrt{\ga}\,\xi\fr{\dot{\phi}^6}{N^6}\bra{2\fr{D_iD^iN}{N}-7\fr{D_iND^iN}{N^2}}\approx 0. \label{Ham_const}
    \ee
As a consistency check, one can obtain an equation which is equivalent to this Hamiltonian constraint in the Lagrangian formalism~(i.e., equivalent to $\delta {\mathcal L}/\delta N$ in the unitary gauge). 
Indeed, without imposing the unitary gauge, on denoting the EOM for the metric as
    \be
    \mE_\mn\coloneqq \fr{1}{\sqrt{-g}}\fr{\delta S}{\delta g^\mn}=0,
    \ee
one can verify that the Hamiltonian constraint is equivalent to $n^\mu n^\nu \mE_\mn=0$, which is, once more, nothing but the time-time component of the metric EOM in the unitary gauge.
Going back to equation \eqref{Ham_const}, for $\xi=0$, the lapse function~$N$ is fixed algebraically from $\pa\mH_N/\pa N\approx 0$. 
On the other hand, for $\xi\ne 0$, \eqref{Ham_const} is an elliptic differential equation for $N$, and hence an appropriate boundary condition should be imposed to fix $N$.
Therefore, the lapse function~$N$ contains the shadowy mode.
It should be noted that the structure of \eqref{Ham_const} is similar to that of \eqref{linear-Poisson}, except that \eqref{Ham_const} is a nonlinear differential equation.
This is as expected, since $\alpha$, which we identified as the shadowy mode in the context of cosmological perturbations, is nothing but the perturbation of the lapse function~$N$.

We can now infer the covariant definition of the shadowy mode.
As $X=-\dot{\phi}^2/N^2$ in the unitary gauge, we expect that $X$ should contain the shadowy mode.
In order to rewrite the Hamiltonian in terms of $X$ instead of $N$, we perform the following canonical transformation:
    \be
    (N,\pi_N)\quad \to \quad
    (X,\pi_X)=\bra{-\fr{\dot{\phi}^2}{N^2},\fr{N^3}{2\dot{\phi}^2}\pi_N},
    \ee
so that we have $\brc{X,\pi_X}_{\rm P}=1$.
Then, the Hamiltonian constraint~\eqref{Ham_const} can be recast as
    \begin{align}
        \xi X^3\bra{\fr{D_iD^iX}{X}+\fr{D_iXD^iX}{4X^2}}+2XP_{X}-P-\frac{3\ti{Q}^2}{4\Mpl^2}
        +\frac{3X\ti{Q}\ti{Q}_X}{\Mpl^2}
        +\frac{\ti{Q}-2X\ti{Q}_X}{\Mpl^2}\frac{\pi}{\sqrt\gamma}\qquad& \nonumber \\
        =\fr{\Mpl^2}{2}R^{(3)}-\fr{1}{\Mpl^2}\fr{2\pi_{ij}\pi^{ij}-\pi^2}{\ga}&. \label{HC1}
    \end{align}
We further make a canonical transformation to replace $X\to Y\coloneqq (-X)^{5/4}$, which brings the above equation in the form of a nonlinear Poisson equation,
    \be
    \xi \Delta Y+f(Y;\ga_{ij},\pi^{ij},R^{(3)})=0, \label{nonlinear-Poisson}
    \ee
where we have denoted $\Delta \coloneqq D_iD^i$ and
    \be
    f=-\fr{5}{4}Y^{-7/5}\brb{\fr{5}{2}YP_Y-P-\fr{3\ti{Q}}{\Mpl^2}\bra{\ti{Q}-\fr{5}{4}Y\ti{Q}_Y}+\fr{1}{\Mpl^2}\fr{\pi}{\sqrt{\gamma}}\bra{\ti{Q}-\fr{5}{2}Y\ti{Q}_Y}-\fr{\Mpl^2}{2}R^{(3)}+\fr{1}{\Mpl^2}\fr{2\pi_{ij}\pi^{ij}-\pi^2}{\ga}}.
    \ee
Note that, even when matter fields exist, the Hamiltonian constraint takes this form and the energy density of the matter fields appears in a generalization of the function~$f$.
It is also possible to find a four-dimensional covariant expression for the Laplacian operator via the St{\"u}ckelberg trick as follows:
    \be
    \xi\Delta Y=\xi\ga^{ij}D_iD_jY\quad\to\quad
    \xi h^{\mn}\na_\mu(h_\nu{}^\lambda\na_\lambda Y),
    \ee
where $h_\mn$ was defined in (\ref{projectiontensor}). Written explicitly,
    \be
    \xi\Delta Y\quad\to\quad
    \xi\brb{\bra{g^\mn-\fr{1}{X}\phi^\mu\phi^\nu}\na_\mu\na_\nu-\fr{1}{X}\bra{\Box\phi-\fr{1}{2X}\phi^\mu\na_\mu X}\phi^\nu\na_\nu}Y.
    \label{covariant_Laplacian}
    \ee
Thus, we have obtained the differential equation that the shadowy mode~$Y$ satisfies in a general coordinate system.
In the equation, the second derivative acting on $Y$ is projected onto the hypersurface which is orthogonal to the timelike vector~$\phi_\mu$. 
Hence, this is indeed a three-dimensional elliptic differential operator on the spacelike hypersurface.
It is also interesting to mention the relation to the covariant EOM.
As mentioned earlier, the Hamiltonian constraint is equivalent to the $n^\mu n^\nu$-component of the metric EOM, $n^\mu n^\nu \mE_\mn=0$, where the left-hand side has the form,
    \begin{align}
    n^\mu n^\nu \mE_\mn=&\;\xi\brb{\fr{X^3}{2}\bra{g^\mn-\fr{\phi^\mu\phi^\nu}{X}}\na_\mu\na_\nu X+\fr{X}{8}(\phi^\mu\na_\mu X)^2+\fr{X^2}{8}\na_\mu X\na^\mu X-\fr{X^2}{2}\phi^\mu\na_\mu X\Box\phi} \nonumber \\
    &+(\text{terms without $\xi$}).
    \end{align}
Here, upon using $Y=(-X)^{5/4}$, one can verify that the terms with $\xi$ coincide with \eqref{covariant_Laplacian} up to an overall factor.

\section{How to deal with the shadowy mode}\label{sec:shadowy}

In the last two sections, we saw how the parameter~$\xi$, which characterizes the deviation from the DHOST class, affects the structure of the Hamiltonian constraint at both the linear and nonlinear levels.
In particular, in the nonlinear Hamiltonian analysis, the constraint equation has the form of a nonlinear Poisson equation~\eqref{nonlinear-Poisson}, which is more complicated than the linear Poisson equation~\eqref{linear-Poisson} for cosmological perturbations.
In this section, we discuss how to treat the nonlinear Poisson equation.

\subsection{Iterative solution}\label{ssec:iterative}

In \S\ref{sec:nonlinear_Hamiltonian}, we obtained the Hamiltonian constraint in the form of a nonlinear Poisson equation~\eqref{nonlinear-Poisson}, which we reproduce here as
    \be
    \xi \Delta Y+f(Y;\ga_{ij},\pi^{ij},R^{(3)},\rho_{\rm m})=0. \label{nonlinear-Poisson2}
    \ee
Here, the energy density of matter fields~$\rho_{\rm m}$ is taken into account.
In the absence of $\xi$, the equation fixes $Y$ algebraically.
On the other hand, for a nonvanishing $\xi$, it is a nontrivial problem to find a solution for the nonlinear Poisson equation.
Nevertheless, so long as the effect of $\xi$ can be considered to be small enough, one can construct a solution in the form of series expansion with respect to $\xi$, as we shall see below.
The solution is assumed to be of the form
    \be
    Y=Y_0+\xi Y_1+\xi^2 Y_2+\cdots, \label{it_sol}
    \ee
and we find each $Y_n$ ($n=0,1,2,\cdots$) in an iterative manner.

At the zeroth order in $\xi$, \eqref{nonlinear-Poisson2} reduces to
    \be
    f(Y_0;\ga_{ij},\pi^{ij},R^{(3)},\rho_{\rm m})=0,
    \ee
which is no longer a differential equation, and hence $Y_0$ can be fixed algebraically. 
Note that this is a nonlinear equation and yields multiple of solutions in general.
For a while, we suppress the second and subsequent arguments of $f$ for notational simplicity.
We choose one of the solutions such that $f_Y(Y_0)\ne 0$, which is necessary for the series expansion with respect to $\xi$ to be valid.
Let us then proceed to find $Y_1$.
Substituting \eqref{it_sol} into \eqref{nonlinear-Poisson2} and taking the terms that are first order in $\xi$, we have
    \be
    \Delta Y_0+f_Y(Y_0)Y_1=0.
    \ee
So long as $f_Y(Y_0)\ne 0$, this equation can be solved for $Y_1$ to obtain
    \be
    Y_1=-\fr{\Delta Y_0}{f_Y(Y_0)}.
    \ee

Likewise, one can iteratively construct a solution for $Y$.
For instance, one has
    \begin{align}
    Y_2=-\fr{2\Delta Y_1+Y_1^2f_{YY}(Y_0)}{2f_Y(Y_0)}, \qquad
    Y_3=-\fr{6\Delta Y_2+6Y_1Y_2f_{YY}(Y_0)+Y_1^3f_{YYY}(Y_0)}{6f_Y(Y_0)}.
    \end{align}
Hence, provided that the series expansion with respect to $\xi$ converges, one can uniquely construct $Y$ once $Y_0$ is fixed.

\subsection{Boundary condition for global solution}\label{ssec:boundary}

In the previous subsection, we studied an iterative solution to the nonlinear Poisson equation~\eqref{nonlinear-Poisson2}. 
However, in reality, there are nonlinear structures in the Universe, and hence the expansion with respect to $\xi$ may break down near such structures.
Nevertheless, going far away from the nonlinear structures, one can still employ the iterative solution there, which can be used as a good boundary condition for solving the nonlinear Poisson equation within the enclosed region.
The solution constructed in this manner would be unique at least locally in the configuration space, though the global uniqueness does not hold in general.
Note that this prescription applies not only when the inhomogeneities have a compact support but when the inhomogeneities exist everywhere.
For instance, we can accommodate cosmological perturbations with a sufficiently small amplitude far from the nonlinear structures of interest.
Note also that our procedure applies only if the equation for the lapse function~$N$ has a nontrivial physical solution when $\xi=0$ (i.e., in the DHOST limit), which is the case for generic U-DHOST theories.
Otherwise, the expansion in the parameter~$\xi$ does not work.\footnote{Therefore, it does not apply to the case of non-projectable Ho\v{r}ava gravity, for which ignoring terms depending on the spatial derivatives of the lapse function in the action results in inconsistencies~\cite{Henneaux:2009zb}. On the other hand, in the projectable Ho\v{r}ava gravity, a shadowy mode is not present and instead the scalar graviton acts as ``dark matter as integration constant'' at low energy~\cite{Mukohyama:2009mz,Mukohyama:2009tp}.}

So far, we focused only on the Hamiltonian constraint and discussed how the value of $Y$ is fixed for given configurations of $\ga_{ij}$, $\pi^{ij}$, and the matter energy density~$\rho_{\rm m}$.
In practice, one has to take into account the evolution equations for the spatial metric and the matter fields.
For a given set of initial conditions for $(\ga_{ij},\pi^{ij},\rho_{\rm m})$, the value of $Y$ at the initial surface is fixed from the Hamiltonian constraint by the prescription mentioned above.
Then, we can compute the value of $(\ga_{ij},\pi^{ij},\rho_{\rm m})$ at the next time step from the evolution equations, which allows us to fix the value of $Y$ at this time step again from the Hamiltonian constraint.
It should be noted that the Hamiltonian constraint should be solved at each time step to determine the nondynamical variable~$N$ (or $X$ or $Y$), as it is no longer a first-class constraint under a fixed gauge.
By repeating this procedure, one can in principle compute the time evolution of all the variables in concern within the region enclosed by the boundary.

\subsection{\texorpdfstring{$\xi\to 0$ limit}{xi to 0 limit}}

We have discussed how to deal with the problem of nonlinear Poisson equation~\eqref{nonlinear-Poisson2} for a given model~\eqref{model} with fixed $\xi$.
Let us now consider the limit~$\xi\to 0$.
Although it is nontrivial whether the limit~$\xi\to 0$ is well defined for the global solution, one can safely take the limit at least near the boundary where the iterative solution derived in \S\ref{ssec:iterative} is assumed to be valid.
Also, we expect that the region where the iterative solution works would become larger and larger as $\xi$ approaches zero, and ultimately it would cover whole the domain of our interest for $|\xi|$ below some critical value.
If this is the case, the $\xi\to 0$ limit of the global solution is well defined.
There might be some pathological cases where this expectation does not work, but any physically plausible setup, in the regime of validity of the effective field theory, would accommodate a well-defined limit.

\section{Conclusions}\label{sec:conclusions}

In the present paper, we have studied the framework of U-DHOST theories, i.e., higher-order scalar-tensor theories which do not belong to the DHOST class but satisfy the degeneracy condition in the unitary gauge.
We have clarified that the apparent Ostrogradsky (actually shadowy) mode satisfies a three-dimensional elliptic differential equation and hence does not propagate.
It should be emphasized that we have taken into account also the dynamics of the metric, which was ignored in \cite{DeFelice:2018ewo}.
For demonstration purposes, we have focused on the model~\eqref{model}, where the deviation from the DHOST class is characterized by the parameter~$\xi$.
In \S\ref{sec:cosmo}, we have studied cosmological perturbations in this model to show that the constraint equation associated with the lapse perturbation, which corresponds to the Hamiltonian constraint, has the form of a Poisson equation with its principal part proportional to $\xi$.
If an appropriate boundary condition is imposed, the constraint equation can be solved uniquely to fix the lapse perturbation.
In \S\ref{sec:nonlinear_Hamiltonian}, we have performed a nonlinear Hamiltonian analysis in the unitary gauge without specifying the background metric.
We have found that, as was the case for cosmological perturbations, the Hamiltonian constraint can be recast in the form of a nonlinear Poisson equation in the presence of $\xi$, which can be regarded as an equation that the shadowy mode satisfies.
Then in \S\ref{sec:shadowy}, we have discussed how one should deal with the shadowy mode.
We have constructed an iterative solution to the nonlinear Poisson equation, which would be useful (at least) as a boundary condition for solving the system of equations within the region enclosed by the boundary.
To reiterate, our analysis is based on the specific model~\eqref{model}, but we expect that it applies to generic U-DHOST theories.
We believe our prescription would also apply to the shadowy mode in the cuscuton models~\cite{Afshordi:2006ad,Iyonaga:2018vnu,Iyonaga:2020bmm} and minimally modified gravity~\cite{DeFelice_2015,Lin:2017oow,Mukohyama:2019unx,Aoki:2018zcv,DeFelice:2020eju,DeFelice:2020prd}.

Having provided a nonlinear definition of the shadowy mode in the U-DHOST theories, it would be intriguing to investigate their phenomenological aspects.
For instance, since the shadowy mode should affect cosmological perturbations as we saw in \S\ref{sec:cosmo}, there would be nontrivial corrections in, e.g., the effective gravitational constant.
Another important issue is to study black holes in the U-DHOST theories.
To this end, we need to consider black hole solutions with a time-dependent scalar hair~(see, e.g., \cite{Mukohyama:2005rw,Babichev:2013cya,Tretyakova:2017lyg,Babichev:2016fbg,Charmousis:2019vnf,Minamitsuji:2019shy,Takahashi:2020hso}) so that one can take the unitary gauge.
Interestingly, there may exist a universal horizon, within which even the instantaneous mode cannot escape to infinity~\cite{Blas:2011ni}.
It would also be interesting to study perturbations about black hole solutions in the U-DHOST theories, following the works~\cite{Ogawa:2015pea,Takahashi:2015pad,Takahashi:2016dnv,Babichev:2018uiw,Takahashi:2019oxz,deRham:2019gha,Khoury:2020aya,Tomikawa:2021pca,Takahashi:2021bml} in the DHOST theories.


\acknowledgments{
The work of A.D.F.~was supported by Japan Society for the Promotion of Science~(JSPS) Grants-in-Aid for Scientific Research No.~20K03969. 
S.M.'s work was supported in part by JSPS Grants-in-Aid for Scientific Research No.~17H02890, No.~17H06359, and by World Premier International Research Center Initiative, MEXT, Japan. 
The work of K.T.~was supported by JSPS KAKENHI Grant No.~JP21J00695.
}


\bibliographystyle{mybibstyle}
\bibliography{bib}

\end{document}